\def\eqnum#1{\eqno (#1)}
\def\fnote#1{\footnote}

\documentstyle[rep10]{article}
\begin{document}
\pagestyle {empty}
Preprint IAE-6051/11
\par
Moscow       1997
\par
\medskip
\bigskip
\par
\bigskip
\par
\centerline{\bf \large Numerical simulation of gliding reflection
} 
\centerline{\bf \large of X-ray beam from rough surface
}
\bigskip
\par
\centerline{T.A.Bobrova, L.I.Ognev
}
\bigskip
\par
\centerline{Abstract
}
\bigskip 
\par
A new method for investigation of X-ray beam propagation in a rough
narrow dielectrical waveguide is proposed on the basis of the numerical
integration of the quazioptical equation. In calculations a model rough
surface is used with the given  distribution of hights of roughness
and given correlation properties of the surface. The method is free
from the limitations such as infiniteness of the surface length and plane
wave approximation which are nessesary for application of Andronov-
Leontovich method. Our method can be applied to any surface with given
nonhomogeniety and distribution of transitional layer.
\newpage
\pagestyle {plain}
\pagenumbering {arabic}
In the middle of 1970-s the experiments on chanalization of soft 
X-rays  \cite{1}  and middle range X-rays \cite{2} 
in glass tubes and also for filtering of hard part of 
X-ray radiation in bent tubes bundles \cite{3}.  
The workability of such systems was demonstrated for transport of
X-ray radiation. 
Later the interest to this manner of X-ray steering was connected 
with the suggestion to use the samples of many specially bent tubes
for geometrical focusing and concentrating of X-ray beam \cite{4}. 
The experiments were performed for transmission of X-ray beam through 
"gap-less collimator" - micron gap betwean two tightly pressed together 
glass plates  \cite{PTE81, PTE84}. 
The revival of interest to localizing of X-rays in thin glass capillaries 
of changing diameter in the condition of total external reflection
at grazing incidence on smooth surface was connected with the attempts 
of microfocusing in a narrowing tube   \cite{5}  (till  $7 \mu m$)  
or in a narrowing polycapillary system  \cite{6}, 
consisting of a great number of melted-in together thin glass tubes
forming one hexahedral block. The problems of the structure of 
the focus of such system deals with the work \cite{NIM97}.
\par
The channalisation of X-ray radiation in hollow glass tubes is possible 
due to the effect of total external reflection \cite{7}.
The dielectric permeability of substance for electro-magnetic radiation
with the energy higher than binding energy of electrons in atoms 
can be approximately written with the use of plasma frequency
$\omega _{p}$ \cite{4}
\par
$$
\epsilon (\omega ) = 1 - (\omega _{p} /\omega )^{2}.
$$
\noindent Because in X-ray energy range  $\epsilon (\omega ) < 1$
and optical density of glass is smaller than the vacuum (or air) optical
density the effect of total external reflection takes place when
the rays come from outside and for this reason is called total external
reflection (TER).
\par
When X-ray radiation comes on sufficiently smooth surface in the conditions
of TER the radiation penetrates only at small depth of the order of  
$60~\AA $,  that maintains the effective reflection and channalisation
of radiation in hollow glass tubes.
\par
Besides of the absorption the effectiveness of reflection depends
strongly on the scattering of radiation on roughness of the surface.
The most of the results on the influence of roughness on the reflection 
of X-ray radiation from surfaces were obtained in the approximation 
of Andronov and Leontovich \cite{8}.
The revue of the results can be found in  \cite{9}.
The approximation is based on the suggestion that the initial wave
coming on the surface is a plane wave and the scattered wave can be 
found for small perturbations of the boundary between two media. 
That is why the investigation of scattreing of wave coming on absorbing 
surface  at small angles is interesting because in this case the Andronov 
and Leontovich approximation apparentely can not be applied any longer.
\par
\medskip
\section{The derivation of master equation.
}
\medskip
Maxwell equations are  \cite{10}
\par
$$
\hbox{rot} {\bf H} = {\epsilon \over c} {\partial {\bf E}\over
\partial t} + {4\pi \sigma \over c} {\bf E},
$$
$$
\hbox{rot} {\bf E} = - {\mu \over c} {\partial  {\bf H}\over \partial  t},
$$
$$
\hbox{div} {\bf E} = 0,
$$
$$
\hbox{div} {\bf H} = 0.
$$
\par
\noindent We can exclude the vector ${\bf H}$
by applying to the second equation 
\par
\noindent the operation $ rot$
\par
$$
\hbox{rot} \hbox{rot} {\bf E} = 
\hbox{grad} \hbox{div} {\bf E} - \Delta  {\bf E}.
$$
\par
\noindent With account of the third Maxwell equation we get 
wave equation:
\par
$$
\Delta  {\bf E} { - }{\epsilon \mu \over c} {\partial ^{2}{\bf
E}\over \partial  t^{2}} + {4\pi \sigma \mu \over c^{2}} {\partial
 {\bf E}\over \partial  t} = 0.
\eqnum{1}
$$
\noindent Furthermore, represent the vector of the electric field 
in the form
\par
$$
{\bf E} = {{\bf e}\over 2} A(x,y,z) \exp [i(\omega t-kz)] + к.с.
\eqnum{2}
$$
\par
\noindent где $k = \sqrt{\epsilon} _{0}{\omega \over c}$.
(In this case $\epsilon _{0}$ is dielectrical permeability of air,
$\epsilon $- dielectrical permeability of glass.)
It can be shown that the speed of changing of the amplitude
$A(z,x,y)$ in the beam is different for the transversal coordinates
$x, y$  and longitudinal coordinate $z$ \cite{10}. 
The evolution along the direction $z$ is much slower than along
transversal coordinates. Thus after substituting of (2) into wave equation
(1)  we can neglect terms  $\partial ^{2}A/\partial z^{2}$  in comparison
with  $k(\partial A/\partial z)$ and
$\partial ^{2}A/\partial x^{2}$, resulting to so called parabolic equation
of quazioptics:
\par
$$
2ik {\partial  A\over \partial  z} =
\Delta _{\perp }A + k^{2} {\delta \epsilon \over \epsilon _{0}} A
\eqnum{3}
$$
\noindent where the difference between the dielectrical permeabilities 
$\delta \epsilon   = \epsilon -\epsilon _{1}$ depends on
coordinates and includes imaginary part corresponding to absorption 
in the substance.
\par
\medskip
\section{
The method of simulation of rough surface.
}
\par
Under the total external reflection at grazing incidence of X-ray beam
the most important is scattering within the plane of incidence
because the scattering withing the plane of the interface is small 
\cite{9}. First of all it is due to the fact that the wavelength of 
the radiation is much smaller than the characteristic scale of
inhomogeneities. That is why for simulation of scattering of X-rays
at TER conditions it is enough to account for only scattering in the 
plane of incidence within 2-dimensional model. In this case the value
$\delta \epsilon $  in the right side of equation (3) becomes the 
function of coordinates $z$ and $x$ :
\par
$$
\delta \epsilon (x,z) = \left\{  0, \quad x > \xi(z)   \atop  {\epsilon
-\epsilon_{0}, \quad x <\xi (z)}\right\},
$$
\noindent where the function  $\xi (z)$ is the hight of the rough 
surface profile and can be regarded as a random value (see Fig.1(а)).
\par
Stationary random value on the interval $(0, Z)$  can be expanded 
in the Furier specious \cite{11}:
\par
$$
\xi (z) =\sum^{\infty }_{k=1}V_{k}\cos
\omega _{k}z + U_{k}\sin  \omega _{k}z, \quad 0 \le  z \le  Z,
$$
\noindent where $V_{k}$ и $U_{k}$
are random amplitudes of the harmonics $\omega _{k}= k \omega _{1}$,
\par
$$
\omega _{1}= 2\pi /Z_{1}
$$
\noindent $Z_{1}$  is maximum spacious period of random sequence. 
In the discrete representation 
\par
$$
\xi (n) =\sum^{m}_{k=0}V_{k}\cos {k\pi n\over N}
+ U_{k}\sin  {k\pi n\over N}, \quad n = \overline {1, N };
$$
\noindent where $V_{k}$ and $U_{k}$ uncorrelated random 
numbers with the dispersion $\sigma ^{2}_{k}$
\par
\noindent and zero mean value. The dispersion of the harmonics is
\par
$$
\sigma ^{2}_{k} = {2\over \pi } \int^{\infty }_{0} R(\xi )
\cos (k\omega _{1}\xi )d\xi ,
\eqnum{4}
$$
\noindent where $R(\xi )$ is correlation function of the random value 
$\xi (z)$. 
For normal random processes the amlitudes $V_{k}$ and  $U_{k}$  
must have normal distribution  \cite{11}. The expression for $\xi (n)$  
can be also repesented as
\par
$$
\xi (n) =\sum^{m}_{k=0}E_{k}\cos  ({k\pi n\over N} + \alpha _{k}),
\eqnum{5}$$
\noindent where $E_{k}$ is random coefficient with Rayleigh 
distribution with the parameter  $\sigma $  equal to $\sigma _{k}$,
where $\alpha _{k}$   is random phase of the harmonics with the
uniform distribution on the interval $(0, 2\pi )$.
\par
For choosing of the number of harmonics  $m$  a relationship can be used
\par
$$
1 - {1\over R(0)} \sum^{m}_{k=0} \sigma ^{2}_{k} \ll  1,
$$
\noindent so the summ of dispersions $\sigma ^{2}_{k}$ 
must be equal to the dispersion of the simulated process.
\par
\medskip
\section{
Numerical results and discussions.
}
The numerical method for solving of the equation (1) was used earlier
for study of motion of channeled electrons in single crystals  \cite{12}
and reflection of positrons from slanting cut single crystals \cite{13}. 
The method implies step by step calculation of the amplitude of the
X-ray wave $A(x,z)$ along the surface of reflection 
starting from its initial value at $z$=0.
\par
For simulation of a rough surface the code was constructed calculating
random sequence of numbers   $\xi (n)$ in accordence with representation
(5). With the given correlation function by relationship (4) 
the amplitudes $E_{k}$ of spectral components were determined.
For simulation of random phase $\alpha _{k}$  random numbers generators
were used. Fragments of random surface  $\xi (n)$  is presented on
on Fig. 1(a). With account of several realisation of numerical
process with different number of points correletion functions were
calculated again along with propability density distribution.
On Fig. 1(b) the comparison of given correlation function (curve 1)
and correlation functions calculated with realisation of $n$=1000 
points (curve 2) and  $n$=10 000 points (curve 3 that nearly coinsides
with 1). It is obvious that  with the increase of $n$  correlation 
function of the process $\xi (n)$  approaches the given function.
It was also shown that the distribution of the propability density 
for random process  $\xi (n)$ approaches Gaussian distribution when
the number of points in the realisation increases (Fig.1(c)).
\par
So we can affirm that the created code simulates random surface 
with the given statistical characteristics and Gaussian propability 
density distribution.
\par
The calculations were made for the radiation with the energy 10 keV
and for interaction with rough glass surface. The width of the correlation
function was chosen $5 \mu m$. The angle of total external reflection
is
$\varphi _{c} =3\cdot 10^{-3} rad.$
\par
When the angle of incidence of X-ray wave is not zero the surface 
can be regarded as infinite and incident wave as a plane wave.
On Fig.~2 the distribution of the intensity of the radiation along the 
coordinate  $x$ (the axis  $x$ is perpendicular to the surface)  
for reflection from smooth surface for the angle of incidence 
$\varphi =10^{-3} rad.$ 
The range $ 0 < x < 2100 \AA $
corresponds to the substance layer.
The oscillations of the intensity are caused by the interference of 
incident and mirror reflected waves.
At some distance from the surface that is defined by numerical
scheme parameters the amplitude was smoothly truncated that is 
dependent on the use of the Furier transformation on the 
$x$ coordinate for solving the equation (3) (see Fig.2).
\par
With the increasing of roughness amplitude the angular spectrum 
of the reflected beam the mirror reflected beam decreases and
at the same time numerous random maxima arise so that clear
interference diffraction picture disappears near the surface.
Simple estimate for roughness tolerance of total external reflection
observation was given in  \cite{14}
$$
h \le  \lambda _{0} / 8 \varphi _{0},
$$
where $\lambda _{0}$ is the radiation wavelength. 
As a clear illustration of inverse dependence of tolerable roughness
amplitude on the incidence angle can be used angular spectra of
reflection shown on Fig. 3.
The spectra were calculated for angles $0.5\cdot 10^{-3} rad.$
(a),$ 10^{-3} rad.$(b), $2\cdot 10^{-3} rad.$(c) 
for the same roughness amplitudes $200 \AA$ 
(here and further the mean squared roughness amplitude is used).
In the first case the influence of roughness on the spectrum is 
small but in the last case the mirror reflected peak can not be 
observed in practice.
\par
The calculations of the distribution of the intensity of radiation
near the smooth reflecting surface can show the depth of penetration 
of X-rays into the surface under the conditions of TER as a
function of the incidence angle and for $\varphi =10^{-3} rad.$ 
it is $60 \AA $,  for  $\varphi =10^{-4}  rad.$ -
$10~\AA $.
For reflection from rough surfaces the depth of penetration
(counted from the mean value of the roughness)
is approximately equal to the mean squared amplitude of roughness.
\par
With the created complex of codes calculations were performed
for transmission of X-ray radiation in a 2-dimentional rough
capillary for zero and near to zero entrance angles.
Transversal dimensions of capillaries were chosen $1 - 2 \mu m$,
that caused diffractional spreading of the plane wave to
the values $0.5\cdot 10^{-4} - 10^{-4} rad.$ 
The length was chosen as $1 - 2 cm$.
\par
On Fig.4 the dependence of integral intensity (over the transversal 
dimension of the capillary) on the distance from the entrance
to the capillary. The angle of incidence $\varphi _{0}$ = 0, 
the width of the capillary is $2 \mu m$,
the roughness is   $1200 \AA $. Abrupt falling of the intensity 
of the radiation in the input of the capillary is connected
with the absorption of the radiation coming on glass butt-ends
and transmitting within the substance.
This part of the radiation can be regarded as fully absorped
at the distance  $z \le  1000 \mu m$ from the input of the
capillary. The analysis of the distribution of losses along the
capillary length for $z \ge  1000 \mu m$ 
reveals its considerable dependence on the distance from the capillary
input. The deeper radiation penetrates the capillary the smaller 
loss normalized  to the length unit is that is apparently related 
to consequent decay of modes the most strongly penetrating into
the substance.
\par
The losses of radiation in the capillary are the greater the
the greater capillary wall roughness is and the bigger the angle 
of incidence of the wave into the capillary is.
The partial dependence of losses of the radiation (
size for zero incidence angle and angle value 
$10^{-4}$ rad is shown on Fig.5. The losses were calculated on the 
range $0.25 cm \le  z \le  1 cm$  from the capillary input.
It is worth noting that losses are not zero even  for the capillary
with smooth walls.
\par
The dependence of losses of radiation in capillaries of various widths
with wall roughness
$800 \AA $  and $400 \AA $ and also without roughness for zero
incidence angle are shown on Fig.6.
The increase of losses when the width of the capillary $\Delta x$
is decreased can be accounted for by the broadening of the plane
wave incident on the input of the capillary
$\Delta \varphi $=$\lambda _{0}$/$\Delta $x.
\par
On Fig.7 radiation angular spectra are shown for capillaries 
with the length 1 cm of various widths with smooth and rough
walls. The calculations were made for zero incidence angle.
The decrease of the square under the spectral peak along with the
increase of roughness can be accounted for by higher absorption 
of the radiation but the shape and the width of the spectral peak
practically does not depend on the roughness size.
\par
\medskip
\section{Conclusion.
}
So the created code can simulate interaction of X-rays with rough
surface (including zero gliding angle) without supposition of
infinite plane wave but with direct account of given surface
relief. That is the advantage of the approach over analytical
methods (incliding Andronov-Leontovich approimation).
In comparison with numerical methods using geometrical optics
the proposed method takes into account the wave nature of the 
radiation (diffraction).
\par
The obtained results enable also to give new interpretation of
the experiments with "gapless collimator" published in 1981-1984
\cite{PTE81, PTE84}. 
Accoding to the data of angular distanse between interference
peaks in transmitted beam with account of diffraction the gap
width is $10 \mu m$ but not $1 \mu m$ as it follows from 
\cite{PTE84} without account for diffraction.
In this case different number of transversal modes is excited
depending on the tilt angle of input beam. On the other hand
if in glass plates unpolished transversal band is left  the 
independence of output beam on the input beam tilt can be explained
due to strong absorption of higher transversal modes along with
retaining of lowest symmetric mode (see Fig.7).
\par

\par
\setlength
\unitlength{1cm}
\newpage
\begin{picture}(11,
14
)
\put(0.5,
14
){\special{em:graph pic1a.gif}}
\put(0,0.1){\parbox[b] {10cm} 
{\footnotesize
Fig.1a. Fragment   of calculated   surface   with  account of
equation (5).
}
}
\end{picture}\\
\newpage
\begin{picture}(11,12)
\put(0.5,12){\special{em:graph pic1b.gif}}
\put(0,0.1){\parbox[b] {10cm} 
{\footnotesize
Fig.1b.   1 - given initially correlation function;
\par
2(dashed)  -  correlation   function   calculated   with
realization of random sequence   (n) with 1000 points,
\par
3 - correlation  function  for  sequence  with  10000
points.
}
}
\end{picture}\\
\newpage
\begin{picture}(11,
14
)
\put(0.5,
14
){\special{em:graph pic1c.gif}}
\put(0,0.1){\parbox[b] {10cm} 
{\footnotesize
Fig.1c.   1 - gaussian distribution with given parameters,
\par
2  -  distribution  of  density  probability  calculated
calculated with realization of random sequence     (n)
with 10000 points.
}
}
\end{picture}\\
\newpage
\begin{picture}(11,
14
)
\put(0.5,
14
)
{\special{em:graph pic2.gif}}
\put(0,0.1){\parbox[b] {10cm} 
{\footnotesize
Fig.2.  Distribution  of  intensity  of $X$-ray  radiation  in  the
vicinity of smooth reflective surface in  the  conditions  of  the
Total External Reflection at the angle of incidence on the surface
\noindent $\varphi _{0} = 10^{-3}$ rad. Axis $x$ is perpendicular 
to the surface.  Reflecting
layer of the substance  is  within $0 < x< 2100 \AA $;  reflecting
surface is marked with dash line.
}
}
\end{picture}\\
\par
\newpage
\begin{picture}(11,
14
)
\put(0.5,
14
){\special{em:graph pic3a.gif}}
\put(0,0.1){\parbox[b] {10cm} 
{\footnotesize
Fig.3. Angular spectrum of reflection of $X$-ray radiation from  the
surface with roughness amplitude $200 \AA $  for incidence angle to the
surface $\varphi _{0} = 0.5\cdot 10^{-3}$ rad.
}
}
\end{picture}\\
\newpage
\begin{picture}(11,
14
)
\put(0.5,14.0){\special{em:graph pic3b.gif}}
\put(0.5,7){\special{em:graph pic3c.gif}}
\put(0,0.1){\parbox[b] {10cm} 
{\footnotesize
Fig.3bc. The same as Fig.3 for $\varphi _{0} = 10^{-3} rad$ (b) and
$\varphi _{0} = 2\cdot 10^{-3} rad$ (c).
}
}
\end{picture}\\
\newpage
\begin{picture}(11,
14
)
\put(0.5,
14
){\special{em:graph pic4.gif}}
\put(0,0.1){\parbox[b] {10cm} 
{\footnotesize
Fig.4. Dependence of integral over  the  width  of  the  capillary
$x~=~2~\mu m$ intensity of radiation on  the  depth  of  capillary  Z.
Angle of incidence of plane wave to the input of the capillary
\noindent $\varphi _{0} = 0,$ wall roughness is $1200 \AA $.
}
}
\end{picture}\\
\newpage
\begin{picture}(11,
10
)
\put(0.5,
10
){\special{em:graph pic5.gif}}
\put(0,0.1){\parbox[b] {10cm} 
{\footnotesize
Fig.5. Dependence of losses of radiation  (\%)  calculated  at  the
distance $0.25 cm < Z < 1.0 cm$ from the input of the  capillary  on
the averaged amplitude of roughness at the capillary 
walls  $(\overline{{\sigma}^{2}_{\xi }})^{1/2}$
- mean squared deviation for roughness $\xi $. Angles of  incidence  of
$X$-ray beams to the input of capillary $\varphi _{0} = 10^{-4}$ rad(1)
\noindent $\varphi _{0} = 0$ rad (2). The capillary width is $2 \mu m$.
}
}
\end{picture}\\
\newpage
\begin{picture}(11,
10
)
\put(0.5,
10
){\special{em:graph pic6.gif}}
\put(0,0.1){\parbox[b] {10cm} 
{\footnotesize
Fig.6. Dependence of losses of radiation (\%) calculated within the
range $0.25 < z < 1.0 cm$  from the input of the capillary. Angle of
incidence $\varphi _{0} = 0.$ 
Roughness of capillaries is $800 \AA  (1), 400 \AA 
(2)$, smooth walls (3).
}
}
\end{picture}\\
\newpage
\begin{picture}(11,
12
)
\put(0.5,
12
){\special{em:graph pic7a.gif}}
\put(0,0.1){\parbox[b] {10cm} 
{\footnotesize
Fig.7a. Angular spectra  of  the  radiation  at  the  output  of
capillaries  with  length $1 cm$  for  smooth  capillaries  (solid
lines) and with rough capillaries $400 \AA  ($dashed lines).  Width  of
capillaries is $0.5 \mu m$. Angle of incidence is $\varphi _{0} = 0.$
}
}
\end{picture}\\
\newpage
\begin{picture}(11,
12
)
\put(0.5,
12
)
{\special{em:graph pic7b.gif}}
\put(0,0.1){\parbox[b] {10cm} 
{\footnotesize
Fig. 7b. The same as Fig.$7a$ but for roughness $800 \AA $ and  width  of
capillaries $1 \mu m$.
}
}
\end{picture}\\
\end{document}